\documentclass[11pt]{revtex4}
\usepackage{amsmath}
\usepackage{amssymb}
\usepackage{amsfonts}
\usepackage{mathrsfs}
\usepackage{epsfig}
\usepackage{bm}
\begin{document}

\title{Effect of the frequency chirp on laser wakefield acceleration}

\author{V. B. Pathak$^{1}$\footnote{vishwa.bandhu@ist.utl.pt}, J. Vieira$^1$, R. A. Fonseca$^{1,2}$, L. O. Silva$^1$\footnote{luis.silva@ist.utl.pt} }
\address{$^1$GoLP/Instituto de Plasmas e Fus\~{a}o Nuclear-Laborat\'orio Associado,  Instituto Superior T\'{e}cnico, 1049-001 Lisboa, Portugal}
\address{$^2$DCTI/ISCTE Lisbon University Institute, 1649-026 Lisbon, Portugal}

\begin{abstract}
The role of laser frequency chirps in the laser wakefield accelerator is examined. We show that in the linear regime, the evolution of the laser pulse length is affected by the frequency chirp, and that positive (negative) chirp compresses (stretches) the laser pulse, thereby increasing (decreasing) the peak vector potential and wakefield amplitude. In the blowout regime, the frequency chirp can be used to fine tune the localized etching rates at the front of the laser. In our simulations, chirped laser pulses can lead to $15 \%$ higher self-trapped electrons, and $10 \%$ higher peak energies as compare to the transform-limited pulse. Chirps may be used to control the phase velocity of the wake, and to relax the self-guiding conditions at the front of the laser. Our predictions are confirmed by multi-dimensional particle-in-cell simulations with OSIRIS.
\end{abstract}

\maketitle


\section {Introduction}
\label{sec: intro}

One of the main goals of plasma based accelerators is to deliver multi-GeV electrons~\cite{bib:blumenfeld_nat_2007, bib:faure_nature_2006, bib:geddes_n_2004, bib:martins_np_2010} in distances that can be orders of magnitude shorter than with standard acceleration techniques. Recent particle-in-cell simulations in Lorentz-boosted frames~\cite{bib:martins_np_2010, bib:martins_pop_2010} predict electron bunch energies beyond $10$ GeV in meter-scale plasmas using next generation $10$ Petawatt laser systems. In fact, accelerating wakefields of nearly $50$ GeV/m have been observed experimentally in plasmas~\cite{bib:blumenfeld_nat_2007}, which are almost $1000$ times higher than the fields observed in conventional accelerators. In laser or plasma wakefield acceleration (LWFA/PWFA), a short laser pulse or ultra relativistic electron beam propagates in an underdense plasma, and excites plasma waves~\cite{bib:tajima_prl_1979, bib:esarey_ieeetranspp_1996} that can trap and accelerate electrons to ultra relativistic energies. This paper examines the role of frequency chirped laser in the LWFA, where the laser pushes plasma electrons away from the propagation axis through the ponderomotive force. This creates a positive space charge, as the ions remain essentially immobile in the time scales associated with the plasma period. Depending upon the laser and plasma parameters, linear or nonlinear plasma waves are excited. If the laser intensity is sufficiently high, the radial ponderomotive force can lead to the cavitation of all the plasma electrons from the region where the laser propagates, creating a spherical plasma wave (bubble or blowout regime~\cite{bib:pukhov_apb_2002,bib:lu_prl_2006}). 

The potential of the blowout regime for several applications has been confirmed in numerous experiments. The large accelerating fields associated with the blowout, which can trap and accelerate plasma electrons (self-injection), lead to the generation of quasi-monoenergetic multi-GeV electrons~\cite{bib:faure_nature_2006, bib:geddes_n_2004, bib:mangles_n_2006, bib:leemans_np_2006, bib:clayton_prl_2010}. In addition, the linear transverse focusing forces associated with the bubble~\cite{bib:lu_prl_2006} are ideal for the generation of X-ray radiation, as the accelerated electron beams perform betatron oscillations in the ion channel~\cite{bib:whittum_prl_1990, bib:wang_prl_2002, bib:kiselev_prl_2004, bib:kneip_n_2010}. The beams obtained from LWFA also have potential to drive a free electron laser~\cite{bib:schoeder_jacow_2006} after solving their present beam quality issues. For these applications it is crucial to control and manipulate the injection process that determines the charge, energy, energy spread, and strength parameter for X-ray radiation~\cite{bib:esarey_pre_2002}. Although several techniques have been proposed to this end, including the use of short plasma down-ramps~\cite{bib:geddes_prl_2008, bib:faure_pop_2010}, the use of transverse external magnetic fields~\cite{bib:jv_prl_2011}, and through the beating structures associated with counter or cross propagating lasers~\cite{bib:faure_nature_2006, bib:kotaki_prl_2009, bib:davoine_prl_2009}, and through ionization mechanisms~\cite{bib:pak_prl_2010}, this paper explores the possibility of using chirped lasers to control self-injection.

Previous investigations on the role of the laser envelope asymmetries in wakefield excitation have already shown that a sharp laser intensity rise can drive stronger wakefields~\cite{bib:bere_pscr_1992, bib:faure_pre_2001, bib:jv_njp_2010}. In addition, theoretical and simulation work~\cite{bib:dodd_pop_2001, bib:schroeder_pop_2003} on the impact of frequency chirps on the long pulse instabilities have shown that the growth of Raman forward scattering like instabilities can be controlled by acting upon the laser frequency chirps. The role of frequency chirps on the laser intensity profile has been explored experimentally~\cite{bib:leemans_prl_2002} in the self-modulated LWFA, showing a significant enhancement in the total charge for sharp rising asymmetric pulses with positive chirps, where the frequency is lower at the front of the laser pulse. In this case, the ponderomotive force is also stronger at the front of the laser, which leads to stronger wakefields. Supported by analytical results, these experiments then emphasized the importance of the asymmetry of the laser pulse on the wake excitation and particle acceleration. 

In this paper we investigate, through numerical simulations with 2D and 3D particle-in-cell (PIC) simulations in OSIRIS~\cite{bib:fonseca_book}, the role of the frequency chirps in the blowout regime to show that the chirp can be used to adjust the self-injection rates, charge, and the output energy of LWFA. For the same plasma density, even though the peak energy of the accelerated electrons are in the same level for the chirped and un-chirped laser pulses, the total injected charge can be increased by $15 \%$, using a chirped pulse laser for these parameters. Positive chirps increase (decrease) the self-injected charge (maximum energy) by up to 10\% in comparison to negative chirps in state-of-the-art conditions. Moreover, frequency chirps also change the laser group velocity in the plasma. These results agree with Ref.~\cite{bib:leemans_prl_2002}, although they are due to a different physical mechanism. 

In Section~\ref{sec: bunch}, an analytical model for the longitudinal bunching of a laser pulse with a frequency chirp is developed in the linear regime. Our model shows that laser with positive (negative) frequency chirp will compress (stretch) throughout the propagation. Accordingly, in the linear regime positive (negative) chirp leads to higher (lower) peak laser intensities. Good agreement between the analytical model and the simulations is found in the linear regime. In Section~\ref{sec: lw}, the evolution of linear wake is investigated for different frequency chirped driving laser pulses. Although initially the wakefield amplitude is nearly independent of the sign of the frequency chirp, at later stages of the laser propagation, the wakefield amplitude increases (decreases) for positively (negatively) chirped pulse. In the blowout regime, which can not be examined by the linear model, the peak intensity increases for positive as well as negative chirps; however, the positively chirped pulse evolves faster than the negatively chirped pulse. In Section~\ref{sec: si} the effect of frequency chirp on self-injection in the blowout regime is analyzed. Finally, the conclusions are stated in Section~\ref{sec: conclusion}.


\section {Longitudinal bunching}
\label{sec: bunch} 

The dominant contributions to the plasma refractive index, associated with the propagation of short laser pulses, are due to the ponderomotive and effective mass nonlinearities. The nonlinear coupling between the laser pulse and the plasma dynamics can lead to several processes, such as laser self-steepening~\cite{bib:jv_njp_2010}, self-compression~\cite{bib:ren_pre_2001}, self-focusing~\cite{bib:mori_ieeeqe_1997}, and self-modulation~\cite{bib:esarey_ieeetranspp_1996, bib:mori_ieeeqe_1997}. For our work, however, the most relevant mechanisms are associated with the self-compression, by which the laser pulse length changes during its propagation in the plasma. 

In order to examine the self-compression of a chirped laser pulse, we will use similar physical arguments to those presented in~\cite{bib:mori_ieeeqe_1997}. Our analysis is valid as long as the envelope approximation can be employed, \textit{i.e.}, as long as $k_0 L_0>>1$, where $k_0$ is the central laser wave number, and $L_0$  is the laser pulse length. This assumption is well verified in the LWFA, where the refractive index for a linearly polarized laser pulse, in the weakly relativistic regime, becomes~\cite{bib:mori_ieeeqe_1997},
\begin{equation}
\label{eq: ri}
\eta=\left[ 1- \frac{\omega_{p}^2}{2 \omega_{0}^2} \left\{ 1+\frac{\delta n}{n_{0}} - \frac{\langle a^2 \rangle} {2} -2  \frac{\delta \omega} {\omega_{0}} \right\} \right],
\end{equation}
and where $\omega_{p} = \sqrt{n_{0}e^{2}/(m\epsilon_{0})}$ is the plasma frequency, $n_{0}$ is the background plasma density, $\omega_{0}$ is the laser frequency, $a= eA/(m_e c^{2})$  is the normalized vector potential, $m_e$ is the electron mass, $c$ is the speed of light, and where $\delta n = n-n_0 $ and $\delta \omega = \omega - \omega_0$ are perturbations to the plasma density and laser central frequency $\omega_0$. In addition, $\langle a^2 \rangle$ is the average of $a^2$ over one laser period. According to the Eq.~(\ref{eq: ri}) the laser group velocity of the laser $v_g=(c^2/v_\phi) /[1+\omega_p^2 (\gamma_0-1)/(2 \omega_0^2\gamma_0 (\gamma_0+1))]$~\cite{bib:mori_ieeeqe_1997, bib:decker_prl_1994}, where $v_\phi =c \eta^{-1}$ is the phase velocity and $\gamma_0=\sqrt{(1+\langle a^2 \rangle/2)}$, can be written as~\cite{bib:mori_ieeeqe_1997}
\begin{equation}
\label{eq: vg}
v_{g}=  c \left[ 1- \frac{\omega_{p}^2}{2 \omega_{0}^2} \left\{1+\frac{\delta n}{n_{0}} - \frac{\langle a^2 \rangle} {4} -2  \frac{\delta \omega} {\omega_{0}}\right\}\right].
 \end{equation}
In Eqs.~(\ref{eq: ri}) and~(\ref{eq: vg}), the second term within the brace ($\{\}$) is related with the plasma density perturbations $\delta n$, the next term represents the relativistic effects associated with the quiver motion of the electrons in the laser field, and the last term is due to the laser frequency modulations. The evolution of the laser pulse length is given by~\cite{bib:mori_ieeeqe_1997}
\begin{equation}
\label{eq: lb1}
\frac{1}{L} \frac{\partial L}{\partial t} =-\frac{1}{c}\frac{\partial v_{g}}{\partial \xi},
\end{equation}
where $L$ is the laser length, $t$ is time, and $\xi = t-z/c$ is the distance in the co-moving frame. Using Eq.~(\ref{eq: vg}), Eq.~(\ref{eq: lb1}) can be rewritten as


\begin{equation}
\label{eq: lb2}
\frac{\partial \log L}{\partial t} =\frac{\omega_{p}^2}{2 \omega_{0}^2} \frac{\partial}{\partial \xi} \left[  \frac{\delta n}{n_{0}} 
                                   - \frac{\langle a^2 \rangle} {4} -2 \frac{\delta \omega} {\omega_{0}}  \right].   
\end{equation}

\begin{figure}[htbp]
\begin{center}
\includegraphics[width=\columnwidth]{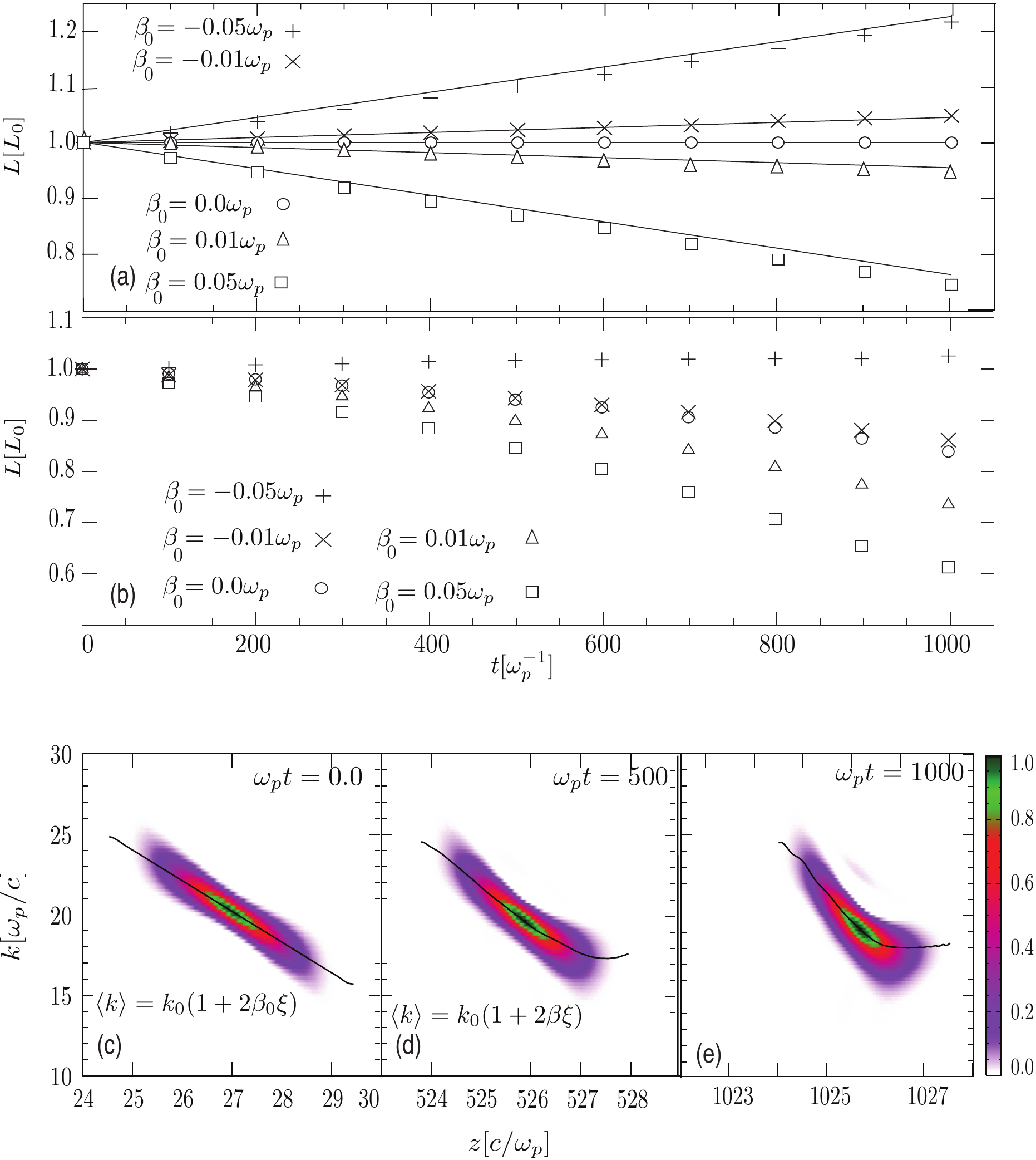}

\caption{\label{fig: fig1} Effect of the initial laser frequency chirp on longitudinal bunching: (a) variation of $L$ (normalized by $L_0$) with time $t$ (normalized by $\omega_{p}^{-1})$ in the linear regime ($a_{0}=0.05$, $\omega_{0}/\omega_{p}=20$, $L_{0}=3c/\omega_{p}$), and (b) $L$ variation with time in the nonlinear regime ($a_{0}=2.0$, $\omega_{0}/\omega_{p}=20$, $L_{0}=5c/\omega_{p}$) for various $\beta_0$, ranging from $-0.01 \omega_p$ to $0.01 \omega_p$. Solid lines in (a) show the theoretical prediction for the pulse length evolution with time for different chirp.  (c), (d), and (e) plot the Wigner transform of the positively chirped laser ($\beta_0=0.05 \omega_p$) used in the linear regime (a), and solid lines represent the $\langle k \rangle$ distribution within the pulse at $\omega_{p} t =0, 500$ and $1000$ respectively.}
\end{center}
\end{figure}

To examine the relative contribution of these terms, we can consider, in the linear regime, the laser pulse profile $ \langle a^2 \rangle \sim a_0^2 \mathrm{sin}^2(c\pi \xi /L)$. For this profile, the density modulation in the wake varies as $\delta n/ n_0 \sim \pi a_0^2 \mathrm{sin}(\omega_p \xi)/8$~\cite{bib:esarey_ieeetranspp_1996}, and the maximum contribution from the first two terms on the right hand side of the Eq.~(\ref{eq: lb2}) is $\sim \omega_p a_0^2/2$. In the regime where $\delta\omega/\omega_0>>\omega_p a_0^2/4$, the effect of ponderomotive and relativistic nonlinearities can be neglected with respect to the term for the spectrum change. In an underdense plasma ($\omega_p^2/\omega_0^2<<1$), for a gaussian chirped laser pulse with electric field $\vec{E}=\vec{E}_0 \exp{(-c^2\xi^2/L^2) \exp(-i \omega_0 (1+\beta \xi)\xi)}$ and chirp coefficient $\beta$,  the average wave number $\langle k\rangle=\int_{-\infty}^{\infty} k \mathcal{W}dk/ (\int_{-\infty}^{\infty} \mathcal{W}dk)$~\cite{bib:silva_pre_1998, bib:serimaa_pra_1986}; where $\mathcal{W}=\int ds \vec{E}(\xi-s/2)\cdot\vec{E}^*(\xi+s/2)\exp(iks)$ is the Wigner transform, can be written as $\langle k \rangle =k_0 (1+2 \beta \xi)$, with $k_0\approx \omega_0$ and $\delta \omega/\omega_0 \approx \delta \langle k \rangle /k_0=2 \beta \xi$. Therefore the first and second term  on the right hand side of Eq.~(\ref{eq: lb2}) can be neglected when $2\beta \xi \sim 2 \beta L >> \omega_p a_0^2/L$. In the remainder of this paper $\beta>>\omega_p a_0^2/2$ will be assumed.

To evaluate further Eq.~(\ref{eq: lb2}) we note that in the absence of any nonlinear plasma effects, the frequency chirp coefficient of a gaussian chirped pulse varies as $\beta=[2 \mathrm{c^2~ln}2/(L^2 \omega_0)]\sqrt{(L/L_{in})^2-1)}$~\cite{bib:silvestri_ieeeqe_1984}, where $L_{in}$ is the pulse length of the transform-limited laser pulse.  As the laser pulse disperses with time, the rate of frequency variation inside the pulse changes, and the frequency chirp coefficient varies as $\beta=\beta_0 (L_0/L)^2 \sqrt{(L^2-L_{in}^2)/(L_0^2-L_{in}^2)}$, where $L_{0}$ and $\beta_0$ are the initial pulse length and the frequency chirp coefficient respectively. For $L$, $L_0>>L_{in}$, frequency chirp coefficient varies as $\beta =\beta_0 L_0/L$. Equation (\ref{eq: lb2}) then yields:
\begin{equation}
\label{eq: pl}
L =  L_{0} \left[\left(1-\frac{L_{in}^2}{L_0^2}\right)\left(1-2 \beta_0 t \frac{\omega_p^2}{\omega_0^2}\frac{1}{1-L_{in}^2/L_0^2}\right)^2+\frac{L_{in}^2}{L_0^2}\right]^{1/2},
\end{equation}
where $L_{in}/L_0 =\sqrt{1/[1+(L_0^2 \omega_0 \beta_0)^2/(2\mathrm{c^2ln2})^2]}$. For the typical parameters of interest for LWFA $(L_0\sim\lambda_p/2,~ \omega_0=20 \omega_p$; $\lambda_p=2\pi c/\omega_p$ is the plasma wavelength), and $\beta_0/\omega_p \sim \mathcal{O}(10^{-2})$, $(L_{in}/L_0)^2 \sim 0.37$ , in this case the expression for the pulse length [Eq.~(\ref{eq: pl})] simplifies to $L =  L_{0} ( 1-2 \beta_0 t \omega_{p}^2/ \omega_{0}^2 )$, which predicts that the positive (negative) frequency chirp compresses (stretches) the laser pulse as it propagates through the plasma. This can be interpreted by investigating the dynamics of different laser photons. For $\beta_0>0$ ($\beta_0<0$), the photons located at the front move slower (faster) than the photons located at the back (front) of the laser. Thus, the distance between the laser photons decreases (increases) as the laser propagates, and laser length thus decreases (increases).
 
In order to further investigate the role of the frequency chirp in the laser pulse length evolution, 1D OSIRIS PIC simulations~\cite{bib:fonseca_book} were performed. The simulation uses a moving window that travels at $c$, with length $30~c/\omega_{p}$, and divided into $3000$ cells, with $500$ particles per cell. The length of the plasma is $1000~c/\omega_{p}$, and the ions form an immobile neutralizing fluid background. In this section the role of the chirp is identified by keeping the laser length, $a_0$, and $\omega_0/\omega_p$ constant regardless of the amount of chirp used. In Section~\ref{sec: si}, studies are shown, where we change the pulse length and the vector potential of the laser consistently with its chirp.

The prediction for longitudinal compression [Eq.~(\ref{eq: pl})]  is in good agreement with the simulations, as illustrated in Fig.~\ref{fig: fig1}. Figure~\ref{fig: fig1}(a) uses a laser with $a_0=0.05$, $\omega_{0}/\omega_{p}=20$ and $L_{0}=3~c/\omega_{p}$, thus exciting linear plasma waves. The laser pulse compresses (stretches) linearly with time for positive (negative) frequency chirp. We note that in these conditions, and in comparison to the scenarios where $\beta_0 \neq 0$, the pulse length remains constant during the laser propagation for $\beta_0 =0.0\omega_p$ (if dispersion effects are neglected). This thus indicates that the ponderomotive and relativistic nonlinearities cancel, confirming that the pulse compression is essentially determined by the initial frequency chirp. Figures~\ref{fig: fig1}(c), (d) and (e) show the Wigner transform~\cite{bib:silva_pre_1998, bib:serimaa_pra_1986}, and average wavenumber distribution ($ \langle k \rangle$) within a chirped laser pulse with $\beta_0 =0.05~\omega_{p}$ at $\omega_{p}t=0$, $\omega_{p}t=500$ and $\omega_{p}t=1000$. The variation in average wavenumber is linear within the pulse until $\omega_p t \sim 500$, and varies as $\langle k \rangle = k_0 (1+2\beta\xi)$, where $\beta= \beta_0 =0.05 \omega_p$ at $\omega_p t =0$, and $\beta\approx \beta_0 (L/L_0)=0.057~\omega_p$ at $\omega_p t=500$. For time $\omega_p t >1000$, $\langle k\rangle$ does not change linearly within the laser pulse. The laser evolution can not be predicted by the theory discussed here for $\omega_p t >1000$ since non-linear mechanism such as self-steepening~\cite{bib:jv_njp_2010} starts to be relevant.

Figure~\ref{fig: fig1}(b) shows the evolution of the laser length using $a_{0}=2.0$ for various frequency chirps varying from $\beta_0 =-0.05~\omega_{p}$ to $0.05~\omega_{p} $. In this particular case, pulse compression is observed for the range of frequency chirps $\beta_0 \geq -0.05~\omega_{p}$; however, the compression was relatively stronger for positively chirped pulses. At higher intensities ($a_{0}\sim1$), in conditions where our analytical model is not valid, the two nonlinearities (ponderomotive and relativistic mass)  do not cancel each other entirely, resulting into the net compression of the pulse~\cite{bib:jv_njp_2010}, and including the contribution of frequency chirp will produce the effects as shown in Fig.~\ref{fig: fig1}(b).
 
In this section we have shown that the initial frequency chirp can compress or stretch the laser pulse depending upon the sign of $\beta_0$. In the next section we examine the effect of frequency chirp on the wakefield excitation in the linear regime due to the longitudinal bunching.

%

 
\section{Wakefield excitation in linear regime}
\label{sec: lw}

The plasma density modulations driven by a linearly polarized laser with normalized vector potential $a_L=a_{0}\times \mathrm{Z}(z,t)\times \cos[{\omega_{0}(1+\beta \xi) \xi}]$ are given by~\cite{bib:esarey_ieeetranspp_1996, bib:keinigs_pf_1987} 

\begin{equation}
\label{eq: weq}
\left( \frac{\partial^{2}}{\partial t^{2}}+\omega_{p}^{2}\right)\frac{\tilde{n}}{n_{0}}=\frac{c^{2}}{2}\frac{\partial^2}{\partial \xi^2} |a_{L}^{2}|
\end{equation}
where, $a_{0}$ is the initial peak normalized vector potential, $\mathrm{Z}(z,t)$ is the longitudinal profile of  $a_{L}$, $\tilde{n}$ is the density modulation in the wake, and $n_{0}$ is the initial homogeneous plasma density. Considering, the laser electric field in 1D is given by $\vec{E}=\hat{x} E_{0} \exp[{-\xi^2/(2 L(t)^2)}]\cos(\omega \xi)$, then $\mathrm{Z}$ is given as
\begin{equation}
\label{eq: vecp}
\mathrm{Z}=\frac{1}{1+\beta \xi}\sqrt{\frac{L_{0}}{L(t)}} \exp[{-\xi^2/(2 L(t)^2)}], 
\end{equation}
where $L(t)$ is given by Eq.~(\ref{eq: pl}). Further assuming that $\beta L/c<<1$, the 1D solution for Eq.~(\ref{eq: weq}) is~\cite{bib:esarey_ieeetranspp_1996, bib:keinigs_pf_1987}\\
\begin{equation}
\label{eq: wd}
\tilde{n}=\frac{1}{2\omega_{p}} \int_{-\infty}^{\xi}\sin\{\omega_{p}( \xi - \xi ')\} \left[ \frac{\partial^2}{\partial \xi^2}|a^2| \right] _{\xi=\xi '}d \xi '.
\end{equation} 
\begin{figure}[htbp]
\begin{center}
\includegraphics[width=\columnwidth]{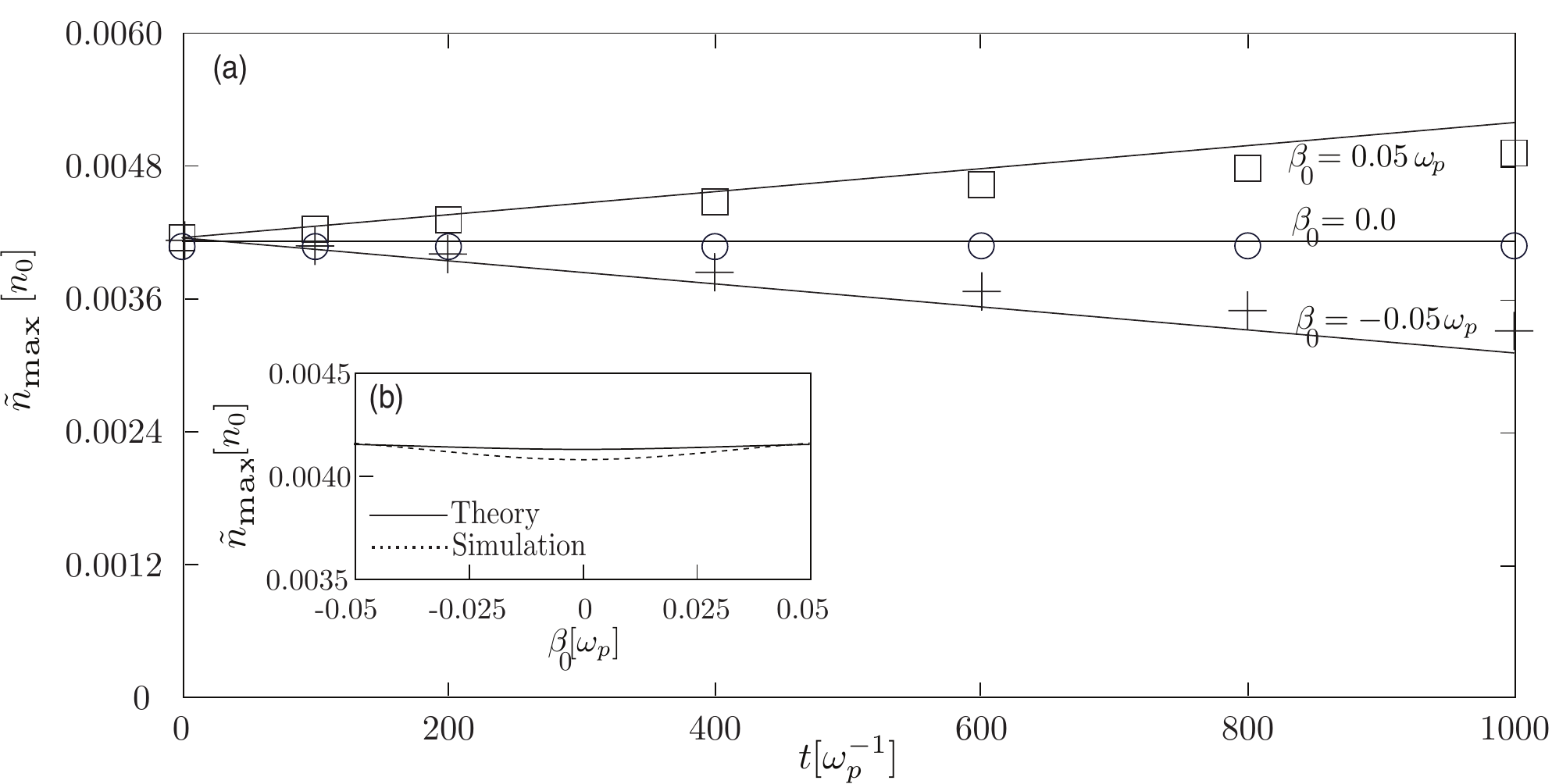}
\caption{\label{fig: fig2} Evolution of peak plasma density ($n_{\mathrm{max}}$) as a function of propagation distance for different chirps and using $a_{0}=0.05$, $\omega_{0}/\omega_{p}=20$, $L_{0}=3c/\omega_{p}$. (a) shows the simulation values of $\tilde{n}_{\mathrm{max}}$ at different times for the chirp coefficients $\beta = 0.0~\omega_p$ by '$\circ$', $-0.05\omega_{p}$ by '$+$', and $0.05\omega_{p}$ by '$\square$'; and (b) compares the theory(dashed line) and simulation(solid line) by plotting the $\tilde{n}_{\mathrm{max}}$ normalized by $n_{0}$ with frequency chirp coefficient $\beta_0$(normalized by $\omega_{p}$) at early stage, neglecting the effect of pump evolution. Solid lines in (a) shows the theoretical predictions.}    
\end{center}
\end{figure}

In order to directly compare the theoretical predictions of the plasma density modulations with the simulations, Eq.~(\ref{eq: wd}) was solved using the same laser pulse profile as used in the simulations. In the simulations we have used a gaussian like 5th order symmetric polynomial profile defined as
\begin{equation}
\label{eq: zsim}
\mathrm{Z}= 10f(z)^3-15 f(z)^4 + 6 f(z)^5, 
\end{equation}
where, for $-L_0<z<0$, $f(z)=(L_{0}+z)/L_{0}$, and for $0<z<L_0$, $f(z)=(L_{0}-z)/L_{0}$. We used the simulation parameters from Section~\ref{sec: bunch}. Figure~\ref{fig: fig2}(b) shows the initial wakefield amplitude for different frequency chirps. For the typical values of $\beta L_0$ used in our simulations, no significant variation in the wake, at initial time, is observed with different $\beta$s, \textit{e.g.}, for $\beta_0 = \pm 0.05~\omega_{p}$, the wakefield amplitude changes by 1$\%$ of the wake amplitude driven by an un-chirped laser pulse. This is because, if we ignore the pulse length variation, the normalized peak laser vector potential is $a_{L}/a_{\beta=0}=1/(1-\beta_0^{2}L_{0}^{2}/c^2)\simeq 1$, being $a_{\beta=0}$ the peak laser vector potential for $\beta_0 = 0$, no significant change in the wakefield with frequency chirp should be expected. Figure~\ref{fig: fig2}(a) demonstrates, however, that the frequency chirp ($\beta_0 \ge 0.05~\omega_p$) can significantly change the amplitude of the plasma wave during the laser propagation. Since the laser pulse compresses or stretches according to the sign of $\beta_0$ (cf. Sec.~\ref{sec: bunch}), increasing or decreasing the laser vector potential, the amplitude of the plasma density modulation also changes accordingly. The evolution of the wakefield amplitude, plotted in Fig.~\ref{fig: fig2}(b), shows that this amplitude increases (decreases) for positively (negatively) chirped pulse, as the laser peak vector potential increases (decreases). Inserting Eq.~(\ref{eq: pl}) in Eq.~(\ref{eq: vecp}), the vector potential can be approximated as 
\begin{equation}
\label{eq:vec_p}
a^{2}\approx a_{0}^{2} (1+2\beta_0 t \omega_{p}^{2}/\omega_{0}^{2}) \mathrm{exp}[-\xi^{2}/L_{0}^{2}]/(1+\beta_0 \xi)^{2}, 
\end{equation}
which, on using in Eq.~(\ref{eq: wd}), gives $\tilde{n} \approx (1+2\beta_0 t \omega_{p}^{2}/\omega_{0}^{2}) \tilde{n}_{0}$, where $\tilde{n}_{0}$ is the initial wake amplitude. Thus,  the density perturbation depends upon the chirp as
\begin{equation}
\label{eq:den_mod}
\tilde{n}/\tilde{n}_{0} - 1 \approx2 \beta_0 \omega_{p}^{2}/\omega_{0}^{2}t.
\end{equation}
The theoretical estimates for the density modulations match with the simulation results as shown in Fig.~\ref{fig: fig2}. 

In a transform limited $20 \mathrm{fs}$ Ti:Sapphire laser pulse, the maximum chirp coefficient that can be introduced is around $\beta_0 \approx1.0 \times 10^{12}~\mathrm{sec^{-2}} \sim 0.01 \omega_p$ by stretching the pulse to $ \sim 28.3 \mathrm{fs}$, for $\omega_0=20 \omega_p$. In such scenarios the wake amplitude changes by only $0.5\%$ at $\omega_p t =1000$. Therefore, pulse compression, and hence the wakefield enhancement due to frequency chirp, play a significant role in LWFA for higher values of frequency chirp coefficients $\beta_0 \ge 0.05 \omega_p$, which can be achieved by stretching a $\sim 12~ \mathrm{fs}$ transform limited laser pulse to a chirped $\sim 18~ \mathrm{fs}$ laser pulse. 

In the blowout regime, where the plasma dynamics is highly nonlinear, effects like self steepening and localized laser absorption also play a significant role in driving a nonlinear wake, and in self-injection. In such scenarios, frequency chirp may influence the laser etching rates, which further can affect the injection rates and beam characteristics. This will be analyzed in the following section.


\section{Blowout regime }
\label{sec: si}

\begin{figure}[htbp]
\begin{center}
\includegraphics[width=\columnwidth]{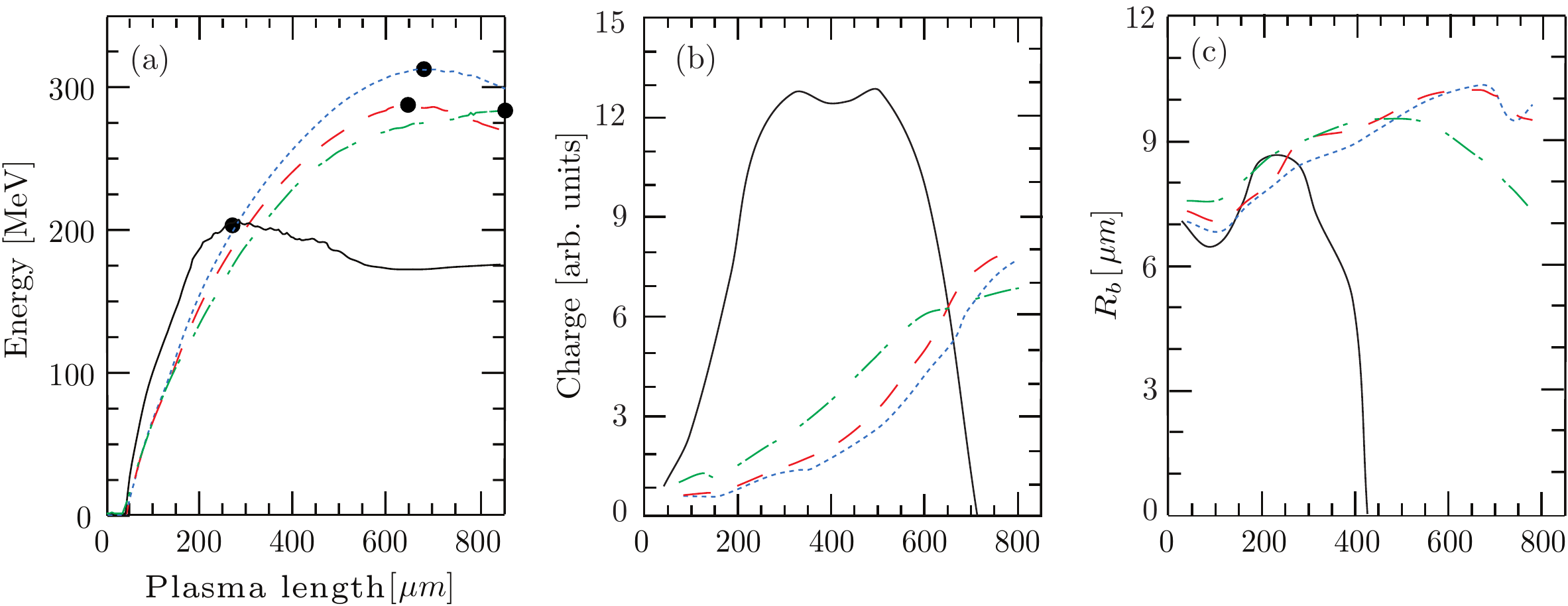}
\caption{\label{fig: fig3} Time evolution of (a) peak energy, (b) total charge injected inside the first bubble, and (c) bubble radius ($R_b$)  in four scenarios (cases). Case (i): ['solid line' (black)] transform-limited pulse with $a_{0}=5.0$, $\omega_{0}=8~\omega_{p}$, $W_0=5~c/\omega_p$ and $L=5~c/\omega_p$. Case (ii): ['dash-dot line' (green)] same transform-limited pulse with $a_{0}=5.0$, $\omega_{0}=11.3~\omega_{p}$, $W_0=3.54~c/\omega_p$ and $L=3.54~c/\omega_p$. Case (iii) ['dashed line' (red)] positively chirped laser pulse with $\beta=0.0055\omega_p$, $a_{0}=4.23$, $\omega_{0}=11.3~\omega_{p}$, $W_0=3.54~c/\omega_p$ and $L=5~c/\omega_p$. Case (iv) ['dotted line' (blue)] negatively chirped pulse with $\beta=-0.0055\omega_p$, $a_{0}=5.0$, $\omega_{0}=11.3~\omega_{p}$, $W_0=3.54~c/\omega_p$ and $L=5~c/\omega_p$. Dark dots in Fig.~\ref{fig: fig3}(a) represents the time at which peak energy, as well as peak efficiency is achieved in LWFA. For cases (i), (iii), and (iv) the efficiencies are the same but greater than the efficiency for the case (ii) by  $\sim 15\%$.}  
\end{center}
\end{figure}

In order to investigate the role of laser frequency chirp in the blowout regime, a set of 3D PIC simulations were performed. For this purpose we consider a linearly polarized 350 mJ transform-limited laser with pulse duration $20~ \mathrm{fs}$ [full width at half maximum of the field(FWHM)], and $ 6~\mathrm{\mu m}$ spot size, with central laser wavelength $\lambda_0=800~$nm. 
In the simulation, these parameters are translated into a transform-limited pulse with $a_{0}=5.0$, with central frequency $\omega_{0}=8~\omega_{p}$, pulse length $L_{0}=5~c/\omega_p$, transverse spot size $W_{0}=5~c/\omega_p$, and with Gaussian transverse field profile given as $\exp(-r^{2}/W_{0}^{2})$, where $r$ is the transverse coordinate, for a homogeneous plasma with plasma-density $n_0=1.75\times10^{19}~\mathrm{cm^{-3}}$. As the pulse is stretched, the maximum chirp coefficient $\beta_{max} = c^2\mathrm{ln2}/L_{in}^2$ can be obtained at $L=1.414~L_{in}$~\cite{bib:silvestri_ieeeqe_1984}.This corresponds to a chirp coefficient of $\beta_0=0.0055~\omega_p$, the peak normalized vector potential is reduced to $a_0=4.23$. Adding this frequency chirp stretches the pulse to $\sim 30~\mathrm{fs}$. To keep the pulse length and plasma wavelength ratio equivalent to the transform-limited laser case, \textit{i.e.}, $L=5~c/\omega_p$, electron density is lowered to $n_0=8.77\times10^{18}~\mathrm{cm^{-3}}$, which translates into $\omega_0=11.3~ \omega_p$, and $W_0=3.54~c/\omega_p$ for a chirped pulse. For a negatively chirped pulse with $\beta=-0.0055~\omega_p$ the rest of the laser simulation parameters are equivalent to the positively chirped pulse ($\beta=0.0055~\omega_p$) case. These chirps can be routinely introduced/controlled in the experiments with lasers~\cite{bib:hong_apb_2002}. In order to compare directly the effect of chirp on the LWFA a simulation using the transform limited pulse propagating in a plasma with density  $n_0= 8.77 \times 10^{18} \mathrm{cm^{-3}}$ (same as used with the chirped pulses) was also performed, using simulation parameters as $a_0=5$, $L_{0}=3.54~c/\omega_p$, $W_{0}=3.54~c/\omega_p$, and $\omega_0=11.3~ \omega_p$. The laser is initialized in a simulation window that moves with the speed of the light, and with dimensions $30c/\omega_{p}\times 36c/\omega_{p} \times 36c/\omega_{p}$, divided into $1800\times180\times180$ cells. The 3D simulations used $2$ particles per cell.

Figure~\ref{fig: fig3} shows the variation in electron beam energy and total charge in the first bucket due to the introduction of frequency chirp in a transform-limited pulse in the 3D simulations. We discuss here results for the four cases, (i) transform-limited pulse with $a_0=5$, $\omega_{0}=8~\omega_{p}$, $L_{0}=5~c/\omega_p$, and $W_{0}=5~c/\omega_p$; (ii) same transform limited pulse (but with plasma density same as used with the chirped pulse) with $a_0=5$, $\omega_{0}=11.3~\omega_{p}$, $L_{0}=3.54~c/\omega_p$, and $W_{0}=3.54~c/\omega_p$; (iii) positively chirped pulse with $\beta=0.0055\omega_p$, $a_0=4.23$, $\omega_{0}=11.3~\omega_{p}$, $L_{0}=5~c/\omega_p$, and $W_{0}=3.54~c/\omega_p$; and (iv) negatively chirped pulse with $\beta=-0.0055\omega_p$, $a_0=4.23$, $\omega_{0}=11.3~\omega_{p}$, $L_{0}=5~c/\omega_p$, and $W_{0}=3.54~c/\omega_p$. Two scenarios are identified. First, when the ratio of pulse duration to plasma wavelength is kept constant by changing the plasma density [Cases (i), (iii) and (iv)]; and second, when plasma density is kept same [Cases (ii), (iii) and (iv)]. In the first scenario, for the chirped pulse [(iii) and (iv)] the peak energy of the accelerated electrons reaches $1.5$ times the peak energy that can be reached using the transform-limited pulse [(i)], however on longer time scales [Fig.~\ref{fig: fig3} (a)]. The higher peak energy for chirped pulse is obtained at the expense of lower total charge [Fig.~\ref{fig: fig3} (b)] as compare to the transform-limited (un-chirped) laser pulse. The main reason behind these differences is the higher $\omega_0/\omega_p$, \textit{i.e.}, lower plasma density in the case of the chirped pulse. Since the laser group velocity ($v_g$), and hence the wake phase velocity $v_{\phi}$, which play a key role in the self-trapping mechanisms in the LWFA~\cite{bib:jv_prl_2011, bib:pak_prl_2010}, is lower for larger densities, the trapping thresholds ~\cite{bib:esarey_ieeetranspp_1996} in case (i) is relaxed as compare to the cases (ii), (iii) and (iv).

In the case of transform-limited laser pulse [case (i)], the peak energy (200 MeV) is achieved after $\sim 200 \mu \mathrm{m}$. At $\sim 400 \mu \mathrm{m}$, the laser becomes pump-depleted, and the transition to the plasma wakefield accelerator (PWFA) was observed. Since the blowout radius is measured at the same longitudinal position in Fig.~\ref{fig: fig3} (c), it drops to zero when the laser pump-depletes. Simulations show that self-injection still occurs when the wake is driven by the laser self-injected electrons, which increases the total charge of the beam at $\sim 500 \mu \mathrm{m}$. Around $\sim 700 \mu \mathrm{m}$, the self-injected electrons can no longer sustain the wake, and are lost to the background plasma. Total acceleration length for LWFA in this case is $\sim 220 \mu m $, which is three times lower as compare to the acceleration lengths observed for the chirped pulse. The dark dots in Fig.~\ref{fig: fig3} (a) represent the points where the peak energy is reached. The efficiency is similar for all the three cases. For the chirped pulse the self-injection and acceleration process is relatively gradual as compare to the un-chirped pulse. The amount of injected charge for the chirped pulse can reach to $60\%$ of the total charge obtained with the transform-limited pulse [case (i)]. 

In the second scenario (keeping plasma density constant), the peak energy for the transform-limited pulse [case (ii)] is $\sim 15 \% $ lower than in the case of negatively chirped pulse [ Fig.~\ref{fig: fig3}(a)], and with total peak charge $\sim 16 \%$ lesser than the charge obtained with the chirped pulse [Fig.~\ref{fig: fig3} (b)]. Hence, using a chirped pulse in place of a transform limited pulse provides better efficiency, as well as in $\sim 25 \%$ shorter acceleration lengths. 
\begin{figure}[htbp]
\begin{center}
\includegraphics[width=\columnwidth]{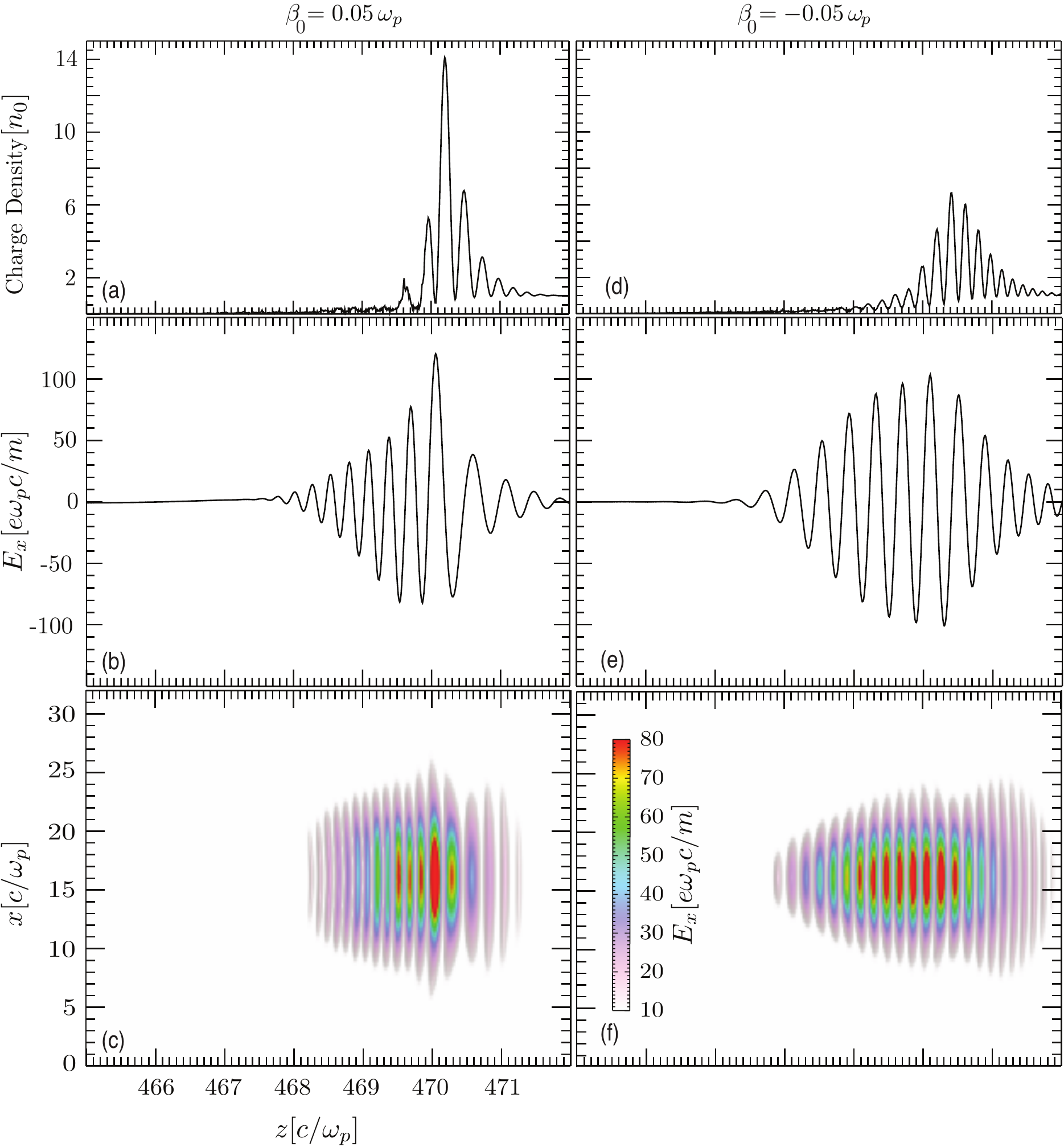}

\caption{\label{fig: fig4} Effect of frequency chirp on localized laser etching: (a) and (d) line out of charge density on the axis at the front of the laser, (b) and (e) line out of laser field on the axis, and (c) and (f) laser field $(E_x)$ in $x-z$ plane at time $448.43 /\omega_{p}$ for $\beta=0.05~\omega_{p}(a, b, c)$ and $\beta=-0.05~\omega_{p} (d, e, f)$ respectively.}  
\end{center}
\end{figure}

Within the chirped pulses, the positively  chirped pulse provides higher (lower) charge (peak energy) as compared to the negatively chirped pulse[Fig.~\ref{fig: fig3}]. Simulations suggested that the higher final electron energies in the case of negatively chirped pulse is due to the fact that negatively chirped laser propagates with higher group velocities, followed by the bubble with higher phase velocity. This increases the threshold for the initial $\gamma$ of the electrons which can be trapped, reducing the number of electrons, as well as electrons that can be accelerated to higher velocities. Beam loading effects~\cite{bib:tzoufras_prl_2008} also play a significant role in reducing the wake field for the positively chirped pulse due to the higher charge self-injection, where as for negatively chirped pulse the beam loading effects are not significant due to relatively lower charge injection as compare to the positively chirped pulse. These simulations showed that by adjusting the initial chirp of the laser pulse, the number of self-injected electrons, and self-injected electron beam energy can be controlled. For the laser parameters discussed here, the difference in total charge for the two chirped cases [(iii) and (iv)] can reach up to $15\%$, and in peak energy $10 \%$ of the difference can be achieved. These results highlights some of the advantages of using chirped pulsed laser in LWFA.

According to the Section ~\ref{sec: lw}, the wake field amplitude is not significantly changed for the frequency chirp discussed here. We believe that it is the combined effect of longitudinal bunching as well as localized etching of the laser which leads to differences in the LWFA by positively and negatively chirped pulse.

Considering the effect of localized etching~\cite{bib:decker_pop_1996}, the laser group velocity $v_g$ is given by $v_g=v_g^{\mathrm{l}}-v_{\mathrm{etch}}$, where $v_{g}^{\mathrm{l}}= c[1 - \omega_p^2/(2 \omega(\xi_\mathrm{etch})^2)]$ is the laser linear group velocity, $v_{\mathrm{etch}}=c\omega_p^2/\omega(\xi_\mathrm{etch})^2$ is the etching velocity~\cite{bib:decker_pop_1996}, and $\xi_\mathrm{etch}$ is the position of the localized etching within the laser in $\xi$ coordinates. A positively chirped pulse, with a red shifted front, etches faster, and may thus propagate with lower group velocity, as compared to a negatively chirped laser pulse. Furthermore, the phase velocity of the plasma wave (wake), propagating at the back of a laser, is roughly equal to the group velocity of the laser; thus, for a positively chirped laser pulse the wake phase velocity is lower as compared to a bubble propagating behind a negatively  chirped laser pulse. Electrons with sufficiently high velocity in the forward direction (this velocity roughly equals the phase velocity of the bubble at the back) have higher probability of getting trapped inside the bubble~\cite{bib:lu_prstab_2007}, which means that the positively chirped laser pulse lowers the threshold for electron trapping by slowing down the wake. On the other hand, a negatively chirped laser pulse, will have a higher group velocity, thus increasing the threshold for trapping.\\

To illustrate clearly the effect of positive and negative frequency chirp on localized etching and on the laser self-guiding~\cite{bib:lu_prstab_2007}, in Fig.~\ref{fig: fig4}, we show the results from 3D PIC simulations for  two linearly polarized laser pulses  with $a_{0}=5.0$, with central frequency $\omega_{0}=20\omega_{p}$, pulse length $L_{0}=3c/\omega_p$, and transverse spot size $W_{0}=4.5 c/\omega_p$, and with $\beta=\pm 0.05\omega_{p}$. We have exaggerated the chirp such that effects attributed to the chirp are clearly visible. Since for $\beta > 0$ the laser intensity is higher at the front than when $\beta<0$, a sharper density spike is formed at the front of the laser for $\beta>0$ [Fig.~\ref{fig: fig4}(a), (d)], which leads to a stronger localized pump-depletion [Fig.~\ref{fig: fig4}(b)-(f)]. For the negatively chirped pulse considered in the simulations, the density spikes are not sharp, so the localized etching is either absent, or very weak, as compared to the positively chirped pulse. Thus, positively chirped lasers may also relax the self-guiding conditions for stable laser propagation in the blowout regime. These results indicate that the variation in laser group velocity and localized pump depletion, due to the initial laser frequency chirp, may impact the rates at which self-injection occurs, and number of self-trapped particles. For the parameters considered in the Fig.~\ref{fig: fig3} [cases (ii) and (iii)], we observe the effect of chirp sign (positive or negative) on localized laser etching; however, and for realistic chirps, the effects are not as strong as shown in Fig.~\ref{fig: fig4}, since the amount of frequency chirp used in the laser for Fig.~\ref{fig: fig4} is 5 times higher as compare to the frequency chirp used in the laser for Fig.~\ref{fig: fig3}. 


\section{Conclusions }
\label{sec: conclusion}
In conclusion, we have examined the effect of the frequency chirp on LWFA. In the linear regime, a positive (negative) frequency chirp compresses (stretches) laser pulse, resulting into enhanced (reduced) wakefield amplitude with time due to dispersive effects. In the blowout regime, and using the chirps that can be routinely introduced in the experiments, simulations show that negatively chirped lasers can provide higher peak energies to the self-injected electrons in comparison to un-chirped lasers. Moreover chirped laser pulses can also lead to higher number of self-injected electrons. In addition, the laser group velocity, and thus the wake phase velocity, is lower (higher) when the laser frequency is lower (higher) at the front of the laser due to positive (negative) frequency linear chirp, which then influences the rate at which self-injection occurs and the number of self-injected electrons, providing an extra control over the self-injection process in LWFA.\\
\\
\textbf{Acknowledgments}

Work partially supported by FCT(Portugal) grants SFRH/BPD/47656/2008, SFRH/BPD/71166/2010 and PTDC/FIS/111720/2009, and by the European Research Council (ERC-2010-AdG Grant 267841). Authors are thankful to Dr. G. Figuira for valuable discussions, and they would also like to acknowledge the assistance of high performance computing resources (Tier-0) provided by PRACE on Jugene based in Germany. Simulations were performed on the Jugene supercomputer (Germany) and on the IST Cluster (IST Lisbon). \\
 \\
 \\
 \textbf{References}\\

\end{document}